\def\rfrs#1#2{(\ref{#1})-(\ref{#2})}
\def\leti{Lense--Thirring}

\def\bar{\begin{eqnarray}}
\def\ear{\end{eqnarray}}

\def\eqi{\begin{equation}}
\def\eqf{\end{equation}}
\def\eqia{\begin{eqnarray}}
\def\eqfa{\end{eqnarray}}

\def\ct#1{\cite{#1}}
\def\lb#1{\label{#1}}

\def\oc2{$\mathcal{O}(c^{-2})$}

\documentclass[11pt]{article}
\usepackage{amsmath,amsthm,amscd,amssymb}
\usepackage{latexsym}
\usepackage{graphicx,epsfig}

\begin{document}

\noindent{\bf \LARGE{On the impossibility of using certain
existing spacecraft for the  measurement of the Lense-Thirring
effect in the terrestrial gravitational field}}
\\
\\
\\
{L. Iorio}\\
{\it Viale Unit$\grave{a}$ di Italia 68, 70125\\Bari, Italy
\\tel./fax 0039 080 5443144
\\e-mail: lorenzo.iorio@libero.it}

\begin{abstract}
In the context of the currently ongoing efforts to improve the
accuracy and reliability of the measurement of the Lense-Thirring
effect in the gravitational field of the Earth it has recently
been proposed to use the data from the existing spacecraft endowed
with some active mechanisms of compensation of the
non-gravitational accelerations like GRACE. In this paper we
critically discuss this interesting possibility. Unfortunately, it
turns out to be unpracticable because of the impact of the
uncancelled even zonal harmonic coefficients  of the multipolar
expansion of the terrestrial gravitational potential and of some
time-dependent tidal perturbations which would resemble as
superimposed linear trends over the necessarily limited
observational time span of the analysis.
\end{abstract}

\section{Introduction}
In this  paper we analyze a recent proposal \ct{fottiti} of using
the data from the existing spacecraft with some active mechanisms
of compensation of the non-gravitational perturbations to improve
the ongoing tests concerning the general relativistic
Lense-Thirring effect \ct{lenti} in the gravitational field of the
Earth.

\section{The proposal of using the existing satellites
endowed with accelerometers}\lb{accel} At the end of Section 5,
 p. 649, right column of \ct{fottiti} it is written: ``[...] one wonders
 why the author of Iorio\footnote{It is the reference \ct{tiinculo} of this paper.} (2005) has not proposed the use of the GRACE satellites which not only have by far
 better shape and orbital stability compared to JASON but they also carry ultra precise accelerometers
 that measure all non--conservative accelerations, so that they are in practice ``free-falling'' particles in vacuum, at least
 to the extent that is covered by the accuracy of these instruments.''

The answer is that the existing spacecraft with some active
mechanism of compensation of the non-gravitational forces (CHAMP,
GRACE, GP-B) fly at too low altitudes in polar orbits, so that the
use of their nodes, combined or not with those of the LAGEOS
satellites, would greatly enhance the systematic errors both due
to the uncancelled even zonal harmonic coefficients $J_{\ell}$ of
the multipolar expansion of the terrestrial gravitational
potential and to certain time-dependent tidal perturbations.
\subsection{The use of the node of GP-B} In
fact, the possible use of the node of the drag-free spacecraft
GP-B has been considered in\footnote{Another analysis can be found
in \ct{pet}.} \ct{gpb}. The major problems come, in this case,
from the fact that a nearly perfect polar orbital configuration
would make the node to precess at a very low rate. This is a
problem because one of the major time-dependent tidal
perturbation, i.e. the solar $K_1$ tide, which is not cancelled
out by the linear combination approach because it is a tesseral
($m=1$) perturbation \ct{iortid}, has just the same period of the
satellite's node. So, over an observational time span of about one
year--which is the lifetime of GP-B--it would resemble an aliasing
superimposed secular trend \ct{pol}. For example, in \ct{gpb} a
combination with the node of LAGEOS and GP-B has been considered:
the coefficient of GP-B would amount to -398, thus fatally
enhancing such tidal bias.
\subsection{The use
of the nodes of the other existing geodetic satellites and of
CHAMP and GRACE} Analogous conclusions can also be drawn for a
possible use of CHAMP and GRACE. Indeed, let us consider, e.g., a
combination with the nodes of LAGEOS, LAGEOS II, Ajisai,
Starlette, Stella, CHAMP and one of the two spacecraft of the
GRACE mission in order to cancel out the first six even zonal
harmonics $J_{\ell},\ell=2,4,6,8,10,12$. Apart from the
coefficient of LAGEOS, which is 1, the coefficients which weigh
the nodes of the other satellites are
\begin{eqnarray}
c_{\rm LAGEOS\ II}&=&0.3237,\\
c_{\rm Ajisai}&=&-0.0228,\\
c_{\rm Starlette}&=&0.0234,\\
c_{\rm Stella}&=&-0.1814,\\
c_{\rm CHAMP}&=&-12.377,\lb{ccha}\\
c_{\rm GRACE}&=&33.5348\lb{cgra}.
\end{eqnarray}
The slope of the gravitomagnetic trend is 3714.58 milliarcseconds
per year (mas yr$^{-1}$ in the following). The systematic error
due to the uncancelled even zonal harmonics $J_{14},\ J_{\rm 16},\
J_{18},...$ amounts to $4.4\%$, at 1-sigma level, according to a
calculation up to degree $\ell=42$ with the variance matrix of the
Earth gravity model EIGEN-CG01C \ct{eigencg01c} which combines
data from CHAMP, GRACE and ground-based measurements. Such error
level is not competitive with those which can be reached by the
combination with LAGEOS and LAGEOS II proposed in\footnote{The
idea of using only the nodes of LAGEOS and LAGEOS II in view of
the expected improvements of our knowledge of the terrestrial
gravitational field with GRACE was put forth for the first time by
Ries et al. in \ct{ries}.} \ct{iormor, iorMGM, iorproc} and used
by Ciufolini, Pavlis and Peron in their recent test \ct{merda},
and the combination involving also Ajisai and Jason-1 \ct{iordorn,
ves}. Indeed, EIGEN-CG01C yields for them a systematic error of
gravitational origin of about $6\%$ and less than  $2\%$,
respectively, at 1-sigma level. Moreover, while the error due to
the geopotential is sensitive to the even zonals, at most, up to
degree $\ell=20$, for such combinations,  the multi-combination
with CHAMP and GRACE is sensitive to a much larger number of even
zonal harmonics due to the inclusion of the lower altitude
satellites Starlette and, especially, Stella, CHAMP and GRACE.
This makes rather difficult and unreliable the evaluation of the
systematic bias induced by the static part of the geopotential
because the calculation of the coefficients $\dot\Omega_{.\ell}$
of the classical secular precessions \ct{cel} yield unstable
results after degree $\ell\sim 40$. Thus, the $4.4\%$ estimate of
the bias due to the even zonals is probably optimistic. Another
serious drawback of such multi-combination is represented by the
relatively large magnitude of the coefficients \rfrs{ccha}{cgra}
which weigh the nodes of CHAMP and GRACE. Indeed, they enhance the
impact of all the uncancelled time-dependent perturbations among
which the solar $K_1$ tide is one of the most powerful.  The
periods of the related orbital perturbations amount to -2.63 years
for CHAMP and -7.23 years for GRACE. Such effects would represent
serious aliasing bias over the necessarily short observational
time span due to the limited lifetimes of CHAMP and GRACE with
respect to the geodetic satellites.
\subsection{The use of the nodes of the other existing geodetic
satellites} In regard to the possibility of only using the other
existing geodetic spherical satellites, mainly Ajisai, Stella and
Starlette due to their long data records available, this problem
has already been tackled in a number of papers, like \ct{jap, imp}
and \ct{casotto}. The fact that the inclusion of the other
satellites in the linear combination scheme is still not
competitive, although the improvements due to the first models
from CHAMP and GRACE, was shown in \ct{iordorn}: the systematic
error of gravitational origin amounts to 31$\%$ with EIGEN-CG01C.
\subsection{The impact of the new Earth gravity models from CHAMP and GRACE on the use of
a polar orbital geometry} The impact of the most recent Earth
gravity models from CHAMP and GRACE on the use of the node of a
single satellite in polar orbit has been discussed in detail in
\ct{ciolon}. By using EIGEN-CG01C, it turns out that for a
semimajor axis of, e.g., 8000 km, quite larger than GP-B, CHAMP
and GRACE, the systematic error due to the full spectrum of the
even zonal harmonics would amount to $25\%$ for an inclination of
88 deg. In this case the period of the $K_1$ tide would amount to
$\sim 10^3$ days. Instead, for an inclination of 89.9 deg the bias
due to the even zonals would be 2$\%$ but the period of the tidal
perturbation would raise to $\sim 10^4$ days. Also with the new
terrestrial gravity field solutions the linear combination
approach would fail.
\section{The proposed use of Jason-1}\lb{jaso}
Iorio proposed to investigate the possibility of analyzing a
suitable linear combination involving the nodes of LAGEOS, LAGEOS
II, Ajisai and Jason-1 in\footnote{See also \ct{ves} on the same
topic. } \ct{iordorn}.

The advantages of the combination involving also Jason-1 are
\begin{itemize}
\item
Cancellation of the first three even zonal harmonics $J_2,\ J_4,\
J_6 $  of the geopotential along with their secular variations
$\dot J_2$, $\dot J_4$, $\dot J_6$. Moreover, the systematic error
due to the remaining higher degree even zonal harmonics $J_8,\
J_{10},...$ is almost model-independent: indeed, it is $\lesssim
2\%$ (1-sigma level) according to the 2nd generation GRACE-only
Earth gravity models EIGEN-GRACE02S \ct{eigengrace02s} $(2\%)$ and
GGM02S $(2.7\%)$ and to the model EIGEN-CG01C \ct{eigencg01c}
$1.6\%)$. It should be mentioned that it is expected that GRACE
will yield a larger improvement in the knowledge of the higher
degree even zonal harmonics, to which the Jason's combination is
sensitive, instead of the lower degree even zonals, which mainly
affect the node-node LAGEOS-LAGEOS II combination. This means that
it may happen that, in the near future, the bias due to the
geopotential will reduce down to $\sim 1\%$ for the Jason's
combination in a satisfactorily model-independent way, while it
may remain more or less unchanged for the two-nodes LAGEOS-LAGEOS
II  combination. The latest results obtained with the combined
model EIGEN-CG03C \ct{cg03} seems to confirm this trend
\ct{iorcul}.

Finally, only the first ten even zonal harmonics (i.e. up to
$J_{20}$) should be accounted for in the sense that the systematic
error due to the geopotential does not change with the inclusion
of more zonals in the calculation.
\item Small coefficient, $\sim 10^{-2}$, of the node of Jason-1.
This is particularly important for reducing the impact of the
non-gravitational acceleration suffered by Jason-1.
\item No secular or long-period perturbations of gravitational and
non-gravitational origin.
\end{itemize}
The possible weak points of this proposals are mainly the
following
\begin{itemize}
\item The huge impact of the non-gravitational forces--mainly atmospheric
drag, direct solar radiation pressure and Earth's albedo--on
Jason-1 which has not a spherical shape, being endowed with solar
panels. Moreover, they should be modelled in a truly dynamical way
in order to avoid to absorb the Lense-Thirring effect as it would
happen in the empirical reduced-dynamic approach adopted so far.
\item The difficulty of getting a smooth long time series of its
orbit also due to the periodical orbital maneuvers. On the other
hand, it should be noted that, due to the Jason's main goal which
is ocean altimetry, only the radial and along-track in-plane
components of its orbit have received major attention up to now.
Also the orbital maneuvers are mainly, although not entirely, in
the orbital plane. Instead, the node is related to the
cross-track, out-of-plane component of the orbit.
\end{itemize}
However, a detailed, although preliminary, evaluation of the
impact of the Jason's non-conservative forces on the entire
combination has been performed in \ct{iordorn}. No secular effects
occur. On the contrary, a relatively high-frequency (the 120-days
period of the $\beta^{'}$ cycle related to the orientation of the
orbital plane with respect to the Sun) of non-gravitational origin
has been found. Its impact on the suggested measurement of the
Lense-Thirring effect has been evaluated to be $\leq 4\%$ over a
2-years time span (without removing such periodic signal).



\section{On some possible misunderstandings concerning various aspects of the measurement
of the Lense-Thirring effect with the LAGEOS satellites.}
Among the six Keplerian orbital elements in terms of which it is
possible to parameterize the orbital motion of a test particle in
the gravitational field of a central body of mass $M$ \ct{roy},
the longitude of the ascending node $\Omega$, the argument of the
pericentre $\omega$ and the mean anomaly $\mathcal{M}$ undergo
secular precessions due to the even zonal harmonics. The general
relativistic gravitomagnetic force only affects $\Omega$ and
$\omega$ with the secular precessions of the Lense-Thirring effect
\ct{lenti}. In principle, $\Omega, \omega$ and $\mathcal{M}$ could
all be used in order to design suitable linear combinations in
order to cancel out the low-degree even zonal harmonics whose
classical precessions are much larger than the Lense-Thirring
signals of interest. However, the mean anomaly cannot be used at
all because it is sensitive to huge non-gravitational
perturbations which especially affect the Keplerian mean motion
$n=\sqrt{GM/a^3}$ via the indirect effects on the semi-major axis
$a$. They have quadratic signature in time.

The title of Section 5, p. 648 of \ct{fottiti} is: ``On the use of
the mean anomaly and on the use of Jason to measure the
Lense-Thirring effect proposed in Iorio (2004) ''. Moreover,  at
the beginning of Section 5, p. 648, right column, first paragraph
of \ct{fottiti} it is written that ``[...] one of the most
profound mistakes and misunderstandings of Iorio\footnote{It is
the reference \ct{tiinculo} of this paper.} (2005) is the proposed
use of the mean anomaly of a satellite to measure the
Lense-Thirring effect (in some previous paper by the same author
the use of the mean anomaly was also explicitly proposed).''


In regard to the mean anomaly, the work quoted by the authors of
\ct{fottiti} in the title of Section 5 of their paper is an old,
unpublished version of the preprint gr-qc/0412057v1. When the
paper \ct{fottiti} was submitted (23 March 2005) and resubmitted
(4 April 2005) the version v3 of gr-qc/0412057, with substantial
revisions, was already available on the Internet since 22 February
2005. In all the versions of such a preprint the mean anomaly is
not mentioned at all. Moreover, no papers by Iorio in which such
an alleged explicit proposal would appear are quoted in the text
of Section 5 of \ct{fottiti}. In fact, in the whole literature
does not exist any published paper by\footnote{The updated list of
the accepted and published papers by Iorio is available on the
Internet at
http://digilander.libero.it/lorri/list$\_$of$\_$publications.htm }
Iorio in which he explicitly put forth the possibility of using
the mean anomaly of the LAGEOS satellites for measuring the
Lense-Thirring effect. On the contrary, the use of the node and
the perigee of the Earth' satellites is always explicitly
mentioned in many papers where the linear approach combination is
exposed and generalized to other situations (see, e.g. pp. 16-17
of \ct{jap}, pp. 5475--5476 of \ct{imp}, p.3 of \ct{COS}, pp.
289--290 of \ct{cel}, p. 5 and p. 8 of \ct{iorMGM}, pp. 4--5 of
\ct{nova}).
%
%
%
Moreover, in Section 2.1.2,
 p. 607, end of the left column of \ct{tiinculo}, after a general description of
 the linear combination approach for
the measurement of the Lense-Thirring effect, it is unambiguously
written ``In general, the orbital elements employed are the nodes
and the perigees and the even zonal harmonics cancelled are the
first N-1 low-degree ones.'' In Section 5, p.649, left column,
second paragraph of \ct{fottiti} it is written: ``A more detailed
work discussing [...] the use of the mean anomaly and other highly
unfeasible proposals by Iorio (2005) will be the subjects of
following paper''. It is, thus, likely that the authors of
\ct{fottiti}, after a careful inspection of the relevant
literature, will experience some difficulties in finding material
for such an announced paper, at least as far as the use of the
mean anomaly for measuring the Lense-Thirring effect is concerned.


Instead, it turns out that one of the authors of \ct{fottiti} used
the semimajor axis, the eccentricity and the mean anomaly of
LAGEOS II in some (not explicitly released) linear combination
with the nodes of LAGEOS and LAGEOS II and the perigee of LAGEOS
II in some previous tests with the EGM96 Earth gravity model
(Figure 4 and pp. 47-48 of \ct{ERE}; Figure 4 and paragraph (b),
pp. 2376-2377 of \ct{KAZ}).


In Section 4, p. 647 of \ct{fottiti} it is claimed that one of the
authors  would have put forth the idea of suitably combining the
Keplerian orbital elements of the LAGEOS-type satellites in order
to disentangle the Lense-Thirring effect from some of the much
larger biasing secular precessions induced by the even zonal
harmonics of the geopotential since the beginning of his studies
on the measurability of the gravitomagnetic effect, when the
LAGEOS II, launched in 1992, did not exist yet. This claim is
supported by quoting from \ct{ciu86, ciu89}:
 ``[...] A solution would be to orbit several
high-altitude, laser-ranged satellites, similar to LAGEOS, to
measure $J_2,$ $J_4$, $J_6$, etc., and one satellite to measure
$\dot{\it \Omega}_{\rm Lense-Thirring }.$''  It seems more likely
that this statement could, at most, mean that one should first use
many satellites to accurately measure the various Earth's even
zonal harmonics with a sufficiently high accuracy and, then,
analyze the node only of one satellite for safely measuring its
gravitomagnetic Lense-Thirring precession. However, there is no
trace at all of the linear combination approach which will be
introduced only in 1996 \ct{ciu96} in the particular case of the
nodes of LAGEOS and LAGEOS II and the perigee of LAGEOS II.
Indeed, the previously quoted statement comes after a discussion
of the impact of $J_2$ and of the higher degree even zonal
harmonics on the possible use of the node only of LAGEOS;
moreover, the rest of the papers \ct{ciu86, ciu89} deals with the
supplementary configuration LAGEOS-LAGEOS X (later LAGEOS-LAGEOS
III/LARES/WEBER-SAT).

The remarks about the triviality of the simple algebraic linear
system of two equations with two unknowns with which the
LAGEOS-LAGEOS II combination is obtained seem to be  a little
inappropriate because they could also be extended to the original
node-node-perigee combination \ct{ciu96} and the related linear
algebraic system of three equations with three unknowns.

In regard to the combination of the nodes of the LAGEOS
satellites, the papers on it are available on the Internet since
April 2003. Moreover, their author, who personally knows the
authors of \ct{fottiti} having collaborated with them for some
years, sent them various e-mails\footnote{They are all available
on request.} between April and September 2003 with his results
attached. In one case (28 March 2003 and 30 March 2003), Ciufolini
asked Iorio to prepare a short table with his results in .doc
format in view of a video-conference with NASA'officials to be
attended in the following days by Ciufolini. A few days before, 26
March 2003, Iorio e-mailed  the .pdf of \ct{iormor} to Ciufolini.
Later, on 7 September 2003, Iorio discussed the impact of the
GGM01C Earth gravity model on the LAGEOS-LAGEOS II combination
with Pavlis.

Thus, we feel that the style and the content of the comment  in
the middle of left column, p. 648 of \ct{fottiti}: ``In
conclusion, all the claims of Iorio (2005) are simply lacking of
any rational basis: above is shown how much work was already
published on this topic before Iorio (2005) even began to produce
any of his paper on this topic and to rediscover some earlier
results [...]. To avoid the misunderstandings of Iorio (2005), it
would have just been a matter of very careful reading the
previously existing literature on this subject!'' sound rather
inappropriate.


\section{Conclusions}
In this paper we demonstrated that the interesting proposal of
using the orbital data from the existing terrestrial spacecraft
endowed with some active mechanism of compensation of the
non-gravitational perturbations like GP-B, CHAMP and GRACE is not
suitable for measuring the Lense-Thirring effect both because of
their too low altitude and of their nearly polar orbital geometry.
Indeed, they would introduce in the resulting linear combinations
much more even zonal harmonics of high degree (more than
$\ell=40$) which would enhance the induced systematic error and
make unreliable its calculation. Moreover, the polar geometry of
GP-B, CHAMP and GRACE would have, as a consequence, that the
coefficients with which they would enter the combinations would be
larger than 1, thus enhancing various uncancelled time-dependent
long-period perturbations like that due to the $K_1$ tide. The new
Earth gravity models from CHAMP and GRACE do not alter this
situation.



\begin{thebibliography}{xxxxx}

\bibitem{fottiti}
Ciufolini, I., and Pavlis, E.C., On the measurement of the
Lense-Thirring effect using the nodes of the LAGEOS satellites, in
reply to ``On the reliability of the so far performed tests for
measuring the Lense-Thirring effect with the LAGEOS satellites''
by L. Iorio, {\it New Astronomy,} {\bf 10}, 636-651, 2005.




\bibitem{lenti}
Lense, J., and  H. Thirring, \"{U}ber den Einfluss der
Eigenrotation der Zentralk{\"{o}}rper auf die Bewegung der
Planeten und Monde nach der Einsteinschen Gravitationstheorie,
{\it Phys. Z.,} {\bf 19}, 156-163, 1918, translated and discussed
by Mashhoon, B., F. W. Hehl, and D. S. Theiss, On the
Gravitational Effects of Rotating Masses: The Thirring-Lense
Papers, {\it Gen. Rel. Grav.,} {\bf 16}, 711-750, 1984. Reprinted
in: Ruffini, R.J., and Sigismondi, C. (eds.), {\it Nonlinear
Gravitodynamics}, (World Scientific, Singapore), 2003. pp.
349--388.






\bibitem{tiinculo}
Iorio, L., On the reliability of the so far performed tests for
measuring the Lense-Thirring effect with the LAGEOS satellites,
{\it New Astronomy,} {\bf 10}, 603-615, 2005.




\bibitem{pet}
Peterson, G.E., {Estimation of the Lense-Thirring precession using
laser-ranged satellites}, {\it CSR-97-1}, Center for Space
Research, The University of Texas at Austin, 1997.



\bibitem{gpb}
Iorio, L., On the impossibility  of using the longitude of the
ascending node of GP-B for measuring the Lense-Thirring effect,
{\it Gen. Rel. Grav.}, {\bf 37}, 1--8, 2005.


\bibitem{iortid}
Iorio, L., Earth tides and Lense-Thirring effect, {\it Celest.
Mech. \& Dyn. Astron.}, {\bf 79}, 201-230, 2001.

\bibitem{pol}
Iorio, L., A critical approach to the concept of a polar,
low-altitude LARES satellite, {\it Class. Quantum Grav.}, {\bf
19}, L175--L183, 2002.


\bibitem{eigencg01c}
Reigber, Ch., Schwintzer, P., Stubenvoll, R., Schmidt, R.,
Flechtner, F.,  Meyer, U., K\"{o}nig, R., Neumayer, H.,
F\"{o}rste, Ch., Barthelmes, F., Zhu, S.Y., Balmino, G., Biancale,
R., Lemoine, J.-M., Meixner, H., and Raimondo, J.C., A High
Resolution Global Gravity Field Model Combining CHAMP and GRACE
Satellite Mission and Surface Gravity Data: EIGEN-CG01C, {\it J.
of Geodesy}, at press, 2005.

\bibitem{ries}
Ries, J.C., Eanes, R.J., Tapley, B.D., and Peterson, G.E.,
Prospects for an Improved Lense-Thirring Test with SLR and the
GRACE Gravity Mission, in: Noomen, R., Klosko, S., Noll, C., and
Pearlman, M. (eds.), {\it Proc. 13th Int. Laser Ranging Workshop,
NASA CP 2003-212248},  (NASA Goddard, Greenbelt), 2003. (Preprint
http://cddisa.gsfc.nasa.gov/lw13/lw$\_${proceedings}.html$\#$science).


\bibitem{iormor}
Iorio, L., and Morea, A., The impact of the new Earth gravity
models on the measurement of the Lense-Thirring effect, {\it Gen.
Rel. Grav.}, {\bf 36}, 1321--1333, 2004. (Preprint
http://www.arxiv.org/abs/gr-qc/0304011).


\bibitem{iorMGM}
Iorio, L., The new Earth gravity models and the measurement of the
Lense-Thirring effect. Paper presented at the {\it Tenth Marcel
Grossmann Meeting on General Relativity Rio de Janeiro, July
 20-26}, 2003. In: Novello, M., Perez-Bergliaffa, S.
 and Ruffini, R.J. (eds.) {\it Proceedings of the Tenth Marcel Grossmann Meeting on General
 Relativity,} (World Scientific, Singapore), 2005, in press. (Preprint
http://www.arxiv.org/abs/gr-qc/0308022).


\bibitem{iorproc}
Iorio, L., The impact of the new CHAMP and GRACE Earth gravity
models on the measurement of the general relativistic
Lense--Thirring effect with the LAGEOS and LAGEOS II satellites,
in: Reigber, Ch., L\"{u}hr, H., Schwintzer, P., and Wickert, J.
(eds.), {\it Earth Observation with CHAMP - Results from Three
Years in Orbit}, (Springer, Berlin), 2005. pp. 187--192. (Preprint
http://www.arxiv.org/abs/gr-qc/0309092).


\bibitem{merda} Ciufolini, I., and Pavlis, E.C., A confirmation of
the general relativistic prediction of the Lense–-Thirring effect,
{\it Nature}, {\bf 431}, 958--960, 2004. (Submitted 2 June 2004).


\bibitem{iordorn}
Iorio, L., and Doornbos, E., How to reach a few percent level in
determining the Lense-Thirring effect? {\it Gen. Rel. Grav.}, {\bf
37 }, 1059-1074, 2005. (Preprint
http://www.arxiv.org/abs/gr-qc/0404062).


\bibitem{ciolon}
Iorio, L., The impact of the new Earth gravity models on the
measurement of the Lense-Thirring effect with a new satellite,
{\it New Astronomy,} {\bf 10}, 616-635, 2005.




\bibitem{ves}
Vespe, F., and Rutigliano, P., The improvement of the Earth
gravity field estimation and its benefits in the atmosphere and
fundamental physics, paper presented at {\it 35th COSPAR
Scientific Assembly Paris, France, 18 - 25 July 2004},
COSPAR04-A-03614, submitted to {\it Adv. Sp. Res.}

\bibitem{cel}
Iorio, L., The impact of the static part of the Earth's gravity
field on some tests of General Relativity with Satellite Laser
Ranging, {\it Celest. Mech. and Dyn. Astron.}, {\bf 86}, 277--294,
2003.

\bibitem{jap}
Iorio, L., Testing General Relativity with LAGEOS, LAGEOS II and
Ajisai Laser-ranged Satellites, {\it J. of the Geodetic Society of
Japan}, {\bf 48}, 13--20, 2002.

\bibitem{imp}
Iorio, L., Is it possible to improve the present LAGEOS-LAGEOS II
Lense-Thirring experiment?, {\em Class. Quantum Grav.}, {\bf 19},
5473--5480, 2002.


\bibitem{casotto}
Casotto, S, Ciufolini, I., Vespe, F., and Bianco, G., Earth
satellites and Gravitomagnetic Field, {\it Il Nuovo Cimento B},
{\bf 105}, 589--599, 1990.


\bibitem{eigengrace02s}
Reigber, Ch., Schmidt, R., Flechtner, F., K\"{o}nig, R., Meyer, U.
Neumayer, K.-H., Schwintzer, P. and Zhu, S.Y., An Earth gravity
field model complete to degree and order 150 from GRACE:
EIGEN-GRACE02S, {\it J. of Geodyn.}, {\bf 39}, 1--10, 2005.


\bibitem{cg03}
F\"{o}rste, C., Flechtner, F.,  Schmidt, R.,  Meyer, U.,
Stubenvoll, R.,  Barthelmes, F.,  K\"{o}nig, R., Neumayer, K.H.
Rothacher, M., Reigber, Ch., Biancale, R., Bruinsma, S., Lemoine,
J.-M., Raimondo, J.C., A New High Resolution Global Gravity Field
Model Derived From Combination of GRACE and CHAMP Mission and
Altimetry/Gravimetry Surface Gravity Data, {\it Poster
g004$\_$EGU05-A-04561.pdf presented at EGU General Assembly 2005,
Vienna, Austria, 24-29, April 2005}.


\bibitem{iorcul}
Iorio, L., The impact of the new Earth gravity model EIGEN-CG03C
on the measurement of the Lense-Thirring effect with some existing
Earth satellites, 2005. (Preprint
http://www.arxiv.org/abs/gr-qc/0505106).


\bibitem{roy}
Roy, A.E. {\it Orbital Motion}, Fourth edition, (Institute of
Physics Publishing, Bristol, 2003).

\bibitem{COS}
Iorio, L., Testing General Relativity With Satellite Laser
Ranging: Recent Developments, {\it 34th COSPAR Scientific Assembly
Houston, TX, 10-19 October 2002}, COSPAR02-A-02121, {\it Adv. Sp.
Res.}, in press., 2002. (Preprint
http://arxiv.org/abs/gr-qc/0210065).


\bibitem{nova}
Iorio, L., Recent Developments in Testing General Relativity with
Space Geodetic Techniques, in: Lynch, T.V., (ed.), {\it Horizons
in World Physics, Vol. 245}, (Nova, Hauppauge), 2004, pp. 1-25.

\bibitem{ERE} Ciufolini, I., Gravitomagnetic phenomena due to spin,
Lense-Thirring effect and its 1995-2000 measurements with Earth
satellites, in: Pascual-S$\acute{\rm a}$nchez, J.F.,
Flor$\acute{\rm \i}$a, L., San Miguel, A., and Vicente, F. (eds.),
{\it Proceedings of the XXIII Spanish Relativity Meeting on
Reference Frames and Gravitomagnetism}, (World Scientific,
Singapore), 2000, pp. 35--52.

\bibitem{KAZ}
Ciufolini, I., The 1995-99 measurements of the Lense-Thirring
effect using laser-ranged satellites, {\it Class. Quantum Grav.},
{\bf 17}, 2369--2380, 2000.


\bibitem{ciu86}
Ciufolini, I.,  Measurement of
\leti\ drag on high-altitude laser ranged artificial satellite,
{\em Phys. Rev. Lett.}, {\bf 56}, 278--281, 1986.


\bibitem{ciu89}
Ciufolini, I., A comprehensive introduction to the LAGEOS
gravitomagnetic experiment: from the importance of the
gravitomagnetic field in physics to preliminary error analysis and
error budget, {\it Int J Mod Phys A}, {\bf 4}, 3083, 1989.

\bibitem{ciu96}
Ciufolini, I.,  On a new method to measure the gravitomagnetic
field using two orbiting satellites, {\em Il Nuovo Cimento A},
{\bf 109}, 1709--1720, 1996.
































\end{thebibliography}
\end{document}